\documentclass[aps,prl,reprint,amsmath,amssymb,amsfonts,showkeys,showpacs,preprintnumbers]{revtex4-1}
\usepackage{bm,graphicx,xcolor,hyperref,rotating,lineno}

\newcommand{\Ec}{E_{\rm c}}

\begin{document}

\title{Invariance of the correlation energy at high density and large dimension in two-electron systems}

\author{Pierre-Fran\c{c}ois Loos}
\email{loos@rsc.anu.edu.au}
\author{Peter M. W. Gill}
\thanks{Corresponding author}
\email{peter.gill@anu.edu.au}
\affiliation{Research School of Chemistry, 
Australian National University, Canberra, 
Australian Capital Territory, 0200, Australia}

\date{\today}

\begin{abstract}
We prove that, 
in the large-dimension limit, 
the high-density correlation energy $\Ec$ 
of two opposite-spin electrons confined in a 
$D$-dimensional space and interacting 
{\em via} a Coulomb potential is given by 
$\Ec \sim -1/(8D^2)$ for any radial confining potential $V(r)$.
This result explains the observed similarity of $\Ec$ in
a variety of two-electron systems in three-dimensional space.
\end{abstract}

\keywords{
correlation energy, two-electron systems, 
high-density limit, large-dimension limit}
\pacs{
31.15.ac, 31.15.ve, 31.15.xp, 
31.15.xr, 31.15.xt}

\maketitle


Understanding and calculating the electronic correlation energy 
is one of the most important and difficult problems 
in molecular physics.
In this pursuit, the study of high-density correlation energy
using perturbation theory
has been particularly profitable,
shedding light on the physically relevant density regime and
providing {\em exact} results
for key systems, such as the uniform electron gas
\cite{GellMann57} and two-electron systems \cite{BetheSalpeter}.
The former is the cornerstone of the most popular 
density functional paradigm 
(the local density approximation) in solid-state 
physics \cite{ParrYang}; the latter provide important
test cases in the development of new 
explicitly correlated methods \cite{Kutzelnigg85,Nakashima07} 
for electronic structure calculations \cite{Helgaker}.
Atomic units are used throughout.

The high-density correlation energy of 
the helium-like ions is obtained 
by expanding both the exact \cite{Hylleraas30} 
and Hartree-Fock (HF) \cite{Linderberg61} 
energies as series in $1/Z$, yielding
\begin{gather}
\label{Eex}
\begin{split}
	E(Z,D,V) 
	& = E^{(0)}(D,V) Z^2
	+ E^{(1)}(D,V) Z
	\\
	& + E^{(2)}(D,V) 
	+ \frac{E^{(3)}(D,V)}{Z} 
	+ \ldots,
\end{split}
\\
\label{EHF}
\begin{split}
	E_{\rm HF}(Z,D,V) 
	& = E^{(0)}(D,V) Z^2
	+ E^{(1)}(D,V) Z
	\\
	& + E_{\rm HF}^{(2)}(D,V)
	+ \frac{E_{\rm HF}^{(3)}(D,V)}{Z}
	+ \ldots,
\end{split}
\end{gather}
where $Z$ is the nuclear charge, 
$D$ is the dimension of the space
and $V$ is the external Coulomb potential.
Equations \eqref{Eex} and \eqref{EHF} share 
the same zeroth- and first-order energies 
because the exact and the HF treatment have
the same zeroth-order Hamiltonian.
Thus, in the high-density (large-$Z$) limit, 
the correlation energy is
\begin{equation}
\label{Ec}
\begin{split}
	\Ec^{(2)}(D,V)
	& =  \lim_{Z\to\infty} \Ec(Z,D,V) 
	\\
	& = \lim_{Z\to\infty} \left[ E(Z,D,V)-E_{\rm HF}(Z,D,V) \right]
	\\
	& = E^{(2)}(D,V) - E_{\rm HF}^{(2)}(D,V).
\end{split}
\end{equation}
Despite intensive study \cite{Schwartz62,Baker90},
the coefficient $E^{(2)}(D,V)$ has not yet been reported 
in closed form.
However, the accurate numerical estimate 
\begin{equation}
\label{E2-He-3D}
	E^{(2)} = -0.157\;666\;429\;469\;14
\end{equation}
has been determined for the important $D=3$ case \cite{Baker90}.
Combining \eqref{E2-He-3D} with the exact result \cite{Linderberg61}
\begin{equation}
\label{E2HF-He-3D}
	E_{\rm HF}^{(2)} = \frac{9}{32} \ln \frac{3}{4} - \frac{13}{432}
\end{equation}
yields a value of 
\begin{equation}
	\Ec^{(2)} = -0.046\;663\;253\;999\;48
\end{equation}
for the helium-like ions in 3-dimensional space.

In the large-$D$ limit, the quantum world reduces to a
simpler semi-classical one \cite{Yaffe82} and problems that
defy solution in $D=3$ sometimes become exactly solvable.
In favorable cases, such solutions provide useful insight into
the $D=3$ case and this strategy has been successfully
applied in many fields of physics \cite{Witten80,Yaffe83}.
Indeed, just as one learns something about interacting
systems by studying non-interacting ones and
introducing the interaction perturbatively,
one learns something about $D = 3$ by studying
the large-$D$ case and introducing dimension-reduction perturbatively.

Singularity analysis \cite{Doren87} 
reveals that the energies of two-electron atoms possess 
first- and second-order poles at $D=1$, 
and that the Kato cusp \cite{Kato57,Morgan93} 
is directly responsible for the second-order pole.
In our previous work \cite{EcLimit09,Ball}, we have expanded the correlation 
energy as a series in $1/(D-1)$ but, although this is formally correct if
summed to infinite order, such expansions falsely imply higher-order poles at $D=1$.
For this reason, we now follow Herschbach and Goodson \cite{Herschbach86,Goodson87}, 
and expand both the exact and HF energies as series in $1/D$.
Although various possibilities exist 
for this dimensional expansion
\cite{Doren86,Doren87,Goodson92,Goodson93},
it is convenient to write
\begin{align}
	E^{(2)}(D,V)
	& = \frac{E^{(2,0)}(V)}{D^2}
	+ \frac{E^{(2,1)}(V)}{D^3}
	+ \ldots,
	\label{E2DV}
	\\
	E_{\rm HF}^{(2)}(D,V)
	& = \frac{E_{\rm HF}^{(2,0)}(V)}{D^2}
	+ \frac{E_{\rm HF}^{(2,1)}(V)}{D^3}
	+ \ldots,
	\label{EHF2DV}
	\\
	\Ec^{(2)}(D,V)
	& = \frac{\Ec^{(2,0)}(V)}{D^2}
	+ \frac{\Ec^{(2,1)}(V)}{D^3}
	+ \ldots,
	\label{Ec2DV}
\end{align}
where
\begin{align}
	\Ec^{(2,0)}(V) 
	& = E^{(2,0)}(V) - E_{\rm HF}^{(2,0)}(V),
	\\
	\Ec^{(2,1)}(V) 
	& = E^{(2,1)}(V) - E_{\rm HF}^{(2,1)}(V).
\end{align}

Such double expansions of the correlation energy 
were originally introduced for the helium-like ions, 
and have lead to accurate estimations of correlation 
\cite{Loeser87a,Loeser87b} and atomic energies \cite{Loeser87c,Kais93}
{\em via} interpolation and renormalization techniques.
Equations \eqref{E2DV}, \eqref{EHF2DV} and \eqref{Ec2DV} 
apply equally to the $^1S$ ground state of any
two-electron system confined by a spherical potential $V(r)$.

For the helium-like ions, it is known 
\cite{Mlodinow81,Herschbach86,Goodson87} that
\begin{align}
	\Ec^{(2,0)}(V) & = - \frac{1}{8},
	&
	\Ec^{(2,1)}(V) & = - \frac{163}{384},
\end{align}
and we have recently found \cite{EcLimit09} that $\Ec^{(2,0)}(V)$
takes the same value in hookium (two electrons in a parabolic well 
\cite{Kestner62,White70,Kais89,Taut93}), spherium
(two electrons on a sphere \cite{Ezra82,Seidl07b,TEOAS1,Quasi09})
and ballium (two electrons in a ball \cite{Thompson02,Thompson05,Ball}).  
In contrast, we found that $\Ec^{(2,1)}(V)$ is $V$-dependent.
The fact that the term $\Ec^{(2,0)}$ is invariant, while
$\Ec^{(2,1)}$ varies with the confinement potential 
allowed us to explain why the high-density correlation energy 
of the previous two-electron systems are similar, 
but not identical, for $D=3$ \cite{EcLimit09,Ball}.
On this basis, we conjectured \cite{EcLimit09} that
\begin{equation}
\label{conjecture}
        \Ec^{(2)}(D,V) 
	\sim - \frac{1}{8D^2} - \frac{C(V)}{D^3}
\end{equation}
holds for \emph{any} spherical confining potential, where
the coefficient $C(V)$ varies slowly with $V(r)$.

In this Letter, we prove that $\Ec^{(2,0)}$ is indeed
universal, and that, in the large-$D$ limit, the high-density 
correlation energy of the $^1S$ ground state of
two electrons is given by \eqref{conjecture}
for any confining potential of the form
\begin{equation}
\label{V}
	V(r) = \text{sgn}(m) r^m v(r),
\end{equation}
where $v(r)$ possesses a Maclaurin series expansion
\begin{equation}
	v(r) = v_0 + v_1 r + v_2 r^2 + \ldots.
\end{equation}

\begin{table}
\caption{
\label{tab:Ec}
$E^{(2,0)}$, $E_{\rm HF}^{(2,0)}$, $\Ec^{(2,0)}$ 
and $\Ec^{(2,1)}$ coefficients for various systems
and $v(r) = 1$.}
\begin{ruledtabular}
\begin{tabular}{lrccccc}
System		&	$m$		&	$-E^{(2,0)}$		&	$-E_{\rm HF}^{(2,0)}$	&	$-\Ec^{(2,0)}$		&	$-\Ec^{(2,1)}$	\\
\hline                                                                          
Helium		&	$-1$		&		5/8		&	1/2			&	1/8			&	0.424479	\\
Airium		&	1		&		7/24		&	1/6			&	1/8			&	0.412767	\\
Hookium		&	2		&		1/4		&	1/8			&	1/8			&	0.433594	\\
Quartium	&	4		&		5/24		&	1/12			&	1/8			&	0.465028	\\
Sextium		&	6		&		3/16		&	1/16			&	1/8			&	0.486771	\\
Ballium		&	$\infty$	&		1/8		&	0			&	1/8			&	0.664063	\\
\end{tabular}
\end{ruledtabular}
\end{table}

In order to prove the conjecture \eqref{conjecture},
we start with the conventional Schr\"odinger equation
\begin{equation}
\label{Sch}
	\Hat{H} \Psi_D = E_D \Psi_D,
\end{equation}
and the general Hamiltonian 
\begin{equation}
\label{H}
	\Hat{H} 
	= - \frac{1}{2} 
	\left( \nabla_1^2 + \nabla_2^2 \right)
	+ Z^{m+2} \left[ V(r_1) + V(r_2) \right]
	+ \frac{1}{r_{12}},
\end{equation}
where $Z$ is the confinement strength and
$r_{12}= \left| \bm{r}_1 - \bm{r}_2 \right|$ 
is the interelectronic distance.
After the Jacobian-weighted transformation 
\begin{gather}
	\Phi_D = \mathcal{J}^{1/2} \Psi_D,
	\\
	\mathcal{J} = r_1^{D-1} r_2^{D-1} \sin^{D-2} \theta,
	\label{Jacobian}
\end{gather}
where $\theta$ is the interelectronic angle,
the Schr\"odinger equation \eqref{Sch} becomes
\begin{equation}
\label{H-Herschbach}
	\left( \Hat{\mathcal{T}} 
	+ \Lambda\,\Hat{\mathcal{U}} 
	+ Z^{m+2} \Hat{\mathcal{V}} 
	+ \Hat{\mathcal{W}} \right) \Phi_D 
	= E_D\,\Phi_D,
\end{equation}
in which, for states with zero total angular momentum,
the kinetic, centrifugal, external and Coulomb operators 
are respectively
\begin{gather}
	-2 \Hat{\mathcal{T}} = 
	\left( \frac{\partial^2}{\partial r_1^2} + \frac{\partial^2}{\partial r_2^2} \right)
	+ \left( \frac{1}{r_1^2} + \frac{1}{r_1^2} \right) 
	\left( \frac{\partial^2}{\partial \theta^2} + \frac{1}{4} \right),
	\\
	\Hat{\mathcal{U}} = 
	\frac{1}{2 \sin^2 \theta} 
	\left( \frac{1}{r_1^2} + \frac{1}{r_1^2} \right),
	\\
	\Hat{\mathcal{V}} =  
	V(r_1) + V(r_2),
	\\
	\Hat{\mathcal{W}} =
	\frac{1}{\sqrt{r_1^2 + r_2^2 -2 r_1 r_2 \cos \theta}},
\end{gather}
and
\begin{equation}
	\Lambda = \frac{(D-2)(D-4)}{4}.
\end{equation}

We now need to recast the Schr\"odinger equation
so that perturbation theory can be applied.  To achieve this,
we successively introduce the scaled quantities
\begin{align}
	r & \rightarrow \frac{\Lambda}{\kappa Z} r,	&
	Z & \rightarrow \frac{Z}{\kappa},
\end{align}
where $\kappa = \Lambda^{\frac{m+1}{m+2}}$, and
introduce the scaled energy
\begin{equation}
	\mathcal{E}_D = \frac{\kappa^2 Z^2}{\Lambda} E_D,
\end{equation}
The Schr\"odinger equation then takes the simple form
\begin{equation}
\label{Hersch-trans}
	\left( \frac{1}{\Lambda} \Hat{\mathcal{T}} 
	+ \Hat{\mathcal{U}} 
	+ \Hat{\mathcal{V}} 
	+ \frac{1}{Z} \Hat{\mathcal{W}} \right) \Phi_D
	= \mathcal{E}_D \Phi_D,
\end{equation}
and it is clear that perturbation theory can
now be used to expand the energy both 
in terms of $Z$ and $\Lambda$.

In the $D=\infty$ limit, the kinetic term vanishes and
classical electrostatics cause the electrons to settle into a
fixed (``Lewis'') structure \cite{Herschbach86}
that minimizes the effective potential
\begin{equation}
\label{X}
	\Hat{\mathcal{X}} 
	= \Hat{\mathcal{U}} + \Hat{\mathcal{V}} 
	+ \frac{1}{Z} \Hat{\mathcal{W}}.
\end{equation}
The minimization conditions are
\begin{gather}
	\frac{\partial \Hat{\mathcal{X}}(r_1,r_2,\theta)}{\partial r_1} =
	\frac{\partial \Hat{\mathcal{X}}(r_1,r_2,\theta)}{\partial r_2} = 0,
	\label{dW-r}
	\\
	\frac{\partial \Hat{\mathcal{X}}(r_1,r_2,\theta)}{\partial \theta} = 0,
	\label{dW-theta}
\end{gather}
and the stability condition implies $m > -2$.
Assuming that the two electrons are equivalent
\footnote{
In the helium atom, for $Z < Z_{\rm crit}$
($Z_{\rm crit} \approx 1.228$ and 
$Z_{\rm crit}^{\rm HF} \approx 0.8839$),
the configuration $r_1 = r_2$ becomes a
saddle point between two minima corresponding to 
non-symmetric configurations \cite{Herschbach86,Goodson87}. 
This is irrelevant, however, in the high-density regime.},
the resulting exact density and energy are
\begin{gather}
	\left|\Phi_{\infty}\right|^2 
	= \delta(r_1 - r_{\infty}) \delta(r_2 - r_{\infty}) \delta(\theta - \theta_{\infty}),
	\\
	\mathcal{E}_{\infty} 
	= \Hat{\mathcal{X}} (r_{\infty},r_{\infty},\theta_{\infty})
	\label{Einf},
\end{gather}
where $\delta$ is the Dirac delta function.
Substituting Taylor expansions of $r_{\infty}$ 
and $\theta_{\infty}$ into \eqref{dW-r} and 
\eqref{dW-theta} yields
\begin{gather}
	r_{\infty} = \alpha + \frac{\alpha^2}{m+2} 
	\left(\frac{1}{2\sqrt{2}} - \Lambda \frac{m+1}{m} \frac{v_1}{v_0} \right) \frac{1}{Z}
	+ \ldots,
	\label{r-eq}
	\\
	\cos \theta_{\infty} = - \frac{\alpha}{4\sqrt{2}} \frac{1}{Z}
	+ \ldots,
	\label{tetha-eq}
\end{gather}
where $\alpha^{-(m+2)} = \text{sgn}(m) m v_0$.
The $m=0$ case requires special attention, and 
is found by taking the $m \to 0$ limit.

For the HF energy, things are simpler.  The HF wave
function is independent of $\theta$, so the only
angular dependence comes from the Jacobian
\eqref{Jacobian}. Moreover, because
\begin{equation}
	\lim_{D\to\infty} \frac{\sin^{D-2} \theta}{\int_0^{\pi} \sin^{D-2} \theta d\theta}
	= \delta\left(\theta - \frac{\pi}{2} \right),
\end{equation}
it follows \cite{Goodson87} that $\theta_{\infty}^{\rm HF} = \pi/2$.
Solving \eqref{dW-r}, one finds that $r_{\infty}^{\rm HF}$ 
and $r_{\infty}$ are equal to second-order in $1/Z$.
Thus, in the large-$D$ limit, the HF density and energy are
\begin{gather}
	\left|\Phi_{\infty}^{\rm HF}\right|^2 
	= \delta(r_1 - r_{\infty}^{\rm HF}) \delta(r_2 - r_{\infty}^{\rm HF}) \delta(\theta - \frac{\pi}{2}),
	\\
	\mathcal{E}_{\infty}^{\rm HF}
	= \Hat{\mathcal{X}} \left(r_{\infty}^{\rm HF},r_{\infty}^{\rm HF},\frac{\pi}{2}\right),
	\label{EinfHF}
\end{gather}
and correlation effects originate entirely from the fact that
$\theta_\infty$ is slightly greater than $\pi/2$ for finite $Z$.

Expanding \eqref{Einf} and \eqref{EinfHF} 
in terms of $Z$ and $D$ yields
\begin{align}
	E^{(2,0)}(V) 
	& = - \frac{1}{8} - \frac{1}{2(m+2)},
	\\
	E_{\rm HF}^{(2,0)}(V) 
	& = - \frac{1}{2(m+2)},
	\label{EHF20}
\end{align}
thus showing that both $E^{(2,0)}$ and $E_{\rm HF}^{(2,0)}$ 
depend on the leading power $m$ of the external potential but not on $v(r)$.

Subtracting these energies yields
\begin{equation}
	\Ec^{(2,0)}(V) = - \frac{1}{8} \label{Ec00},
\end{equation}
and completes the proof that, in the high-density limit, 
the leading coefficient $\Ec^{(2,0)}$ of the large-$D$
expansion of the correlation energy is universal,
{\em i.e.} it does not depend on the external potential $V(r)$.

What is the origin of the constant in Eq.~\eqref{Ec00}?  It comes directly from
the leading coefficient ($1/4\sqrt{2}$) in the $1/Z$ expansion of $\theta_{\infty}$
(Eq. \eqref{tetha-eq}) and, because that is determined {\em via} Eq. \eqref{dW-theta},
it is independent of the external potential $V(r)$.  This reveals that Eq.~\eqref{Ec00}
applies to a pair of electrons in any radial external potential, but not to \emph{anisotropic}
external potentials.

Detailed analysis of $\Ec^{(2,0)}$ shows that it
results from contributions of $+1/8$ and $-1/4$ from
the centrifugal potential $\Hat{\mathcal{U}}$ and the
Coulomb operator $\Hat{\mathcal{W}}$, respectively.
The external potential $\Hat{\mathcal{V}}$, which
contributes identically in the exact and HF treatments, 
does not contribute to the correlation energy.
Kato has made a similar argument \cite{Kato57} to explain the behavior 
of the wave function as $r_{12} \to 0$.
In a $D$-dimensional space, 
the Kato cusp condition is \cite{Morgan93}
\begin{equation}
\label{cusp}
	\left. \frac{\partial \Psi_D}{\partial r_{12}}\right|_{r_{12}=0} 
	= \frac{1}{D-1} \Psi_D(r_{12}=0),
\end{equation}
and arises from the cancelation of the singularities in the Coulomb
operator and the $D$-dependent 
angular part of the kinetic operator \cite{Helgaker}.
These observations suggest a connection between 
the result \eqref{Ec00} and the Kato cusp \eqref{cusp}.
For large but finite $D$, 
the discovery that the Kato cusp plays a key role in the
large-$Z$ limit would not be surprising for, in this limit, the only relevant
information is the behavior \eqref{cusp}
of the wave function near $r_{12} = 0$.

\begin{figure}
\begin{center}
\includegraphics[width=0.48\textwidth]{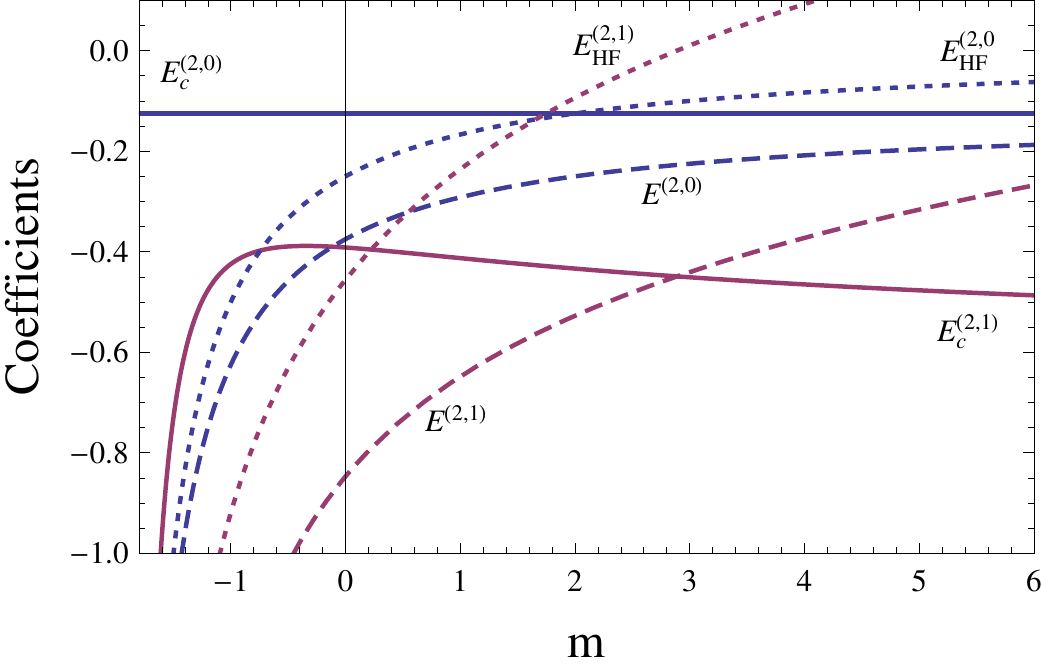}
\caption{\label{fig:res} 
Coefficients of the exact (dashed), HF (dotted) and 
correlation (solid) energies with respect to $m$, 
for $v(r)=1$ (Eqs. \eqref{E2DV}, \eqref{EHF2DV} and \eqref{Ec2DV}).}
\end{center}
\end{figure}

The $E^{(2,1)}$ and $E_{\rm HF}^{(2,1)}$ coefficients
can be found by considering the Langmuir vibrations of the
electrons around their equilibrium positions \cite{Herschbach86,Goodson87}.
The general expressions depend on $v_0$ and $v_1$, 
but are not reported here.  However, for $v(r)=1$, which
includes many of the most common external potentials, we find
\begin{multline}
	\Ec^{(2,1)}(V) = - \frac{85}{128} - \frac{9/32}{(m+2)^{3/2}}\\
	+ \frac{1/2}{(m+2)^{1/2}} + \frac{1/16}{(m+2)^{1/2}+2},
\end{multline}
showing that $\Ec^{(2,1)}$, unlike $\Ec^{(2,0)}$, is potential-dependent.
It is singular at $m=-2$, tends to $-85/128$ as $m \to \infty$,
and reaches a maximum of $-0.388482$ at $m \approx -0.344223$.
The latter value of $m$ corresponds to the minimum 
of the correlation energy in the large-$D$ limit.
Numerical values of $\Ec^{(2,1)}$ are reported in 
Table \ref{tab:Ec} for various systems, and the
components of the correlation energy are shown 
graphically in Fig. \ref{fig:res}.

In conclusion, we have proved that the leading term $\Ec \sim -1/(8D^2)$ in the
large-$D$ expansion of the high-density correlation energy 
of an electron pair is invariant to the nature of the radial confining
potential.  
Although formally divergent \cite{Elout98}, 
truncated $1/D$ expansions have been found to be a powerful tool for the
exploration of correlation effects and, in the present study, they help to 
explain the observation that, in finite-dimensional spaces 
such as $D=3$, the correlation energy depends only weakly 
on the confining potential.

We thank Andrew Gilbert for several stimulating discussions 
at the early stage of this work.
P.M.W.G. thanks the NCI National Facility for a generous grant 
of supercomputer time and the Australian Research Council 
(Grants DP0771978 and DP0984806) for funding.


\end{document}